# Quantum batteries – The future of energy storage?

J. Q. Quach[1,2], G. Cerullo[3,4], T. Virgili[3]

1. CSIRO, Ian Wark Laboratory, Bayview Ave, Clayton, Victoria, 3168, Australia

2. The University of Adelaide, South Australia 5005, Australia

3. Istituto di Fotonica e Nanotecnologia – CNR, IFN - Piazza Leonardo da Vinci 32, 20133 Milano, Italy

4. Dipartimento di Fisica, Politecnico di Milano, Piazza Leonardo da Vinci 32, 20133 Milano, Italy

According to the International Energy Agency, each human uses more than 80 GJ of energy per year; this is equivalent to leaving a washing machine continuously running for one year for every person on Earth. This consumption is expected to increase by 28% by 2040 (from 2015 levels)[1]. The majority (86%) of this energy comes from fossil fuels. This dependence on fossil fuels comes with major environmental costs, with climate change arguably being the greatest challenge facing our era. Renewable energy offers a possible solution. However, renewable energy sources, like solar and wind are not continuous sources, and therefore energy storage technology or batteries, remain an urgent challenge for further worldwide adoption of renewable energy. Alongside the need for efficient batteries to store renewable energy, the portability of batteries makes them an essential component in mobile technologies, including electric vehicles. Current batteries operate on the basis of well-understood electrochemical principles which were developed two centuries ago. While there is an ongoing intense effort aimed at improving their performance through optimization of the materials and the device architecture, it is worth exploring completely novel and disruptive approaches towards energy storage. Quantum batteries are energy storage devices that utilise quantum mechanics to enhance performance or functionality. While they are still in their infancy with only proof-of-principle demonstrations achieved, their radically innovative design principles offer a potential solution to future energy challenges.

**Information to energy**

The importance of quantum technologies and their impact on scientific research and society are growing at an impressive pace. As the 20th century technology has been shaped by electronic and photonic devices, whose operational principles are rooted in quantum physics (the so-called first quantum revolution), the 21st century will be characterized by a completely new class of applications based on our ability to detect and manipulate individual quantum objects (the second quantum revolution). Quantum-mechanical systems present unique characteristics, such as quantum superposition and entanglement, which, besides being conceptually fascinating, enable novel disruptive technological applications. A two-level system (referred to as qubit in the field of quantum computing) is in a quantum superposition state when it is described by the wave function $|\psi\rangle = \alpha |0\rangle + \beta |1\rangle$, with $|\alpha|^2$ and $|\beta|^2$ the probabilities of finding the system in states $|0\rangle$ and $|1\rangle$, respectively. The state of a qubit, like the state of any quantum system, can be described by a vector in a Hilbert space. Multiple quantum systems are said to be entangled when the results of measurements performed on individual systems are correlated in any measurement basis, i.e. the knowledge of the state of one qubit gives the state of the other qubit(s) with 100% certainty. Quantum technologies hold the promise of disruptive conceptual and technological advances in sensing, communications and computation, exploiting the so-called quantum advantages afforded through the science of quantum information. Now there is a deep connection between information and energy, and so one may envisage adopting these quantum advantages in an energy context to develop novel energetic systems, such as quantum batteries, which outperform classical energy storage systems.

The link between information and energy is most colourfully illustrated in a thought experiment known as *Maxwell's demon*[2]. In a simplified version of this thought experiment, one imagines a free particle confined to a box in a thermal bath (Fig. 1). The box is then partitioned with a wall that is free to move. There is an agent (Maxwell's demon) that observes the box. By knowing which side of the partition the particle is in, the demon can attach a weight so that through an adiabatic expansion process, the weight is lifted. As such, one bit of information (the particle is on the left of right side of the partition) is converted to work. This would seem to violate the second law of thermodynamics as work can be cyclically extracted without increasing entropy. However, this is resolved upon realising that for the system to be truly cyclic, the demon's memory needs to be erased as well: this erasure process dissipates heat, increasing entropy.

An interesting phenomenon occurs when one considers the whole system to be quantum mechanical. Here the process of observing and storing the location of the particle entangles the memory with the observed system. This quantum correlation is itself a form of information that can be converted to energy. These correlations underpin the unique properties of quantum batteries.

**Theory – Historical perspective**

Quantum batteries are a redesign of energy storage devices from the bottom up. They are modelled with the simplest quantum energy storage system: a collection of identical qubits, which can be sub-atomic particles, atoms or molecules. In a seminal work, Alicki and Fannes[3] sought to understand whether entanglement could enhance the amount of extractable work in such a model. They demonstrated that entanglement enables a greater amount of extractable work compared to situations where entanglement is absent. Subsequent studies showed that it is possible to diminish entanglement without negatively affecting the maximal work extraction, with the caveat that this reduction in entanglement necessitated more operations or time[4]. This led to the notion that entanglement enhances the rate at which quantum batteries can be charged, as it reduces the number of states required to be traversed between the initial and final states of the associated Hilbert space[4].

Binder et al.[5] supported this hypothesis, demonstrating that *N* entangled spins possess the ability to charge *N* times faster than *N* spins that do not interact. The properties of physical systems normally can be categorised as intensive (i.e. they are independent of the system size, such as density or temperature) or extensive (i.e. they grow in proportion to system size, such as mass or entropy). The charging rate of quantum batteries however is a superextensive property, as it grows exponentially with size: the charging per unit scales with *N*, meaning that quantum batteries with larger capacity actually take less time to charge, a counterintuitive and fascinating behaviour radically different from that of classical batteries.

The studies mentioned so far have assumed *N* separate quantum systems which are globally entangled, which poses significant challenges in practical implementation. Ferraro et al.[6] addressed this issue by demonstrating that, by connecting all spins coherently to a single quantum energy source within a photonic cavity resonant to their transition energy, it becomes possible to achieve effective long-range interactions among all the spins. This approach, referred to as the Dicke quantum battery due to the governing Hamiltonian, showed a charging power that scaled with $\sqrt{N}$ for large *N*. This conceptual advancement significantly improved the feasibility of physically realizing quantum batteries, by simplifying their architecture.

Subsequent research indicated that the acceleration in charging speed of the Dicke quantum battery is not attributed to entanglement but rather to an increased effective coupling strength of the cavity, which emerges from coherent cooperative interactions[7]. This contention between superextensive charging with and without entanglement was resolved by Julia-Farre et al.[8]. Through a geometric analysis of the associated Hilbert space they found that the charging power is bounded by the square root of the product of the quantum Fisher information and the variance of the quantum battery Hamiltonian. The former is related to the speed of the trajectory in the associated energy eigenspace, and the latter encodes non-local correlations. The superextensive charging of the Dicke quantum battery is an example of the former, global entangling operators of the latter. More generally, there are two types of quantum battery properties: those that arise out of the intrinsic quantum nature of the system, and those that come explicitly from quantum many-body interactions.

**Experimental platforms**

From Alicki and Fannes' seminal paper[3], it took another ten years for superextensive charging to be experimentally demonstrated using organic microcavities[9]. In this section we review the crop of experimental quantum battery platforms, which can be classified into room and cold temperature experiments (Fig 2).

*Room-temperature experiments*

The key advantage of room-temperature quantum batteries is that they can perform in less restrictive conditions, as compared to their low-temperature counterparts. Quantum systems are strongly affected by decoherence, i.e. the loss of quantum superposition due to the interaction with the environment, which projects the wave function of the system on one of its eigenstates. As decoherence rapidly increases with temperature, the interplay between coherence and decoherence rates is critical in room temperature experiments. For instance, one requires the energy level spacing in the quantum system to significantly exceed the energy of thermal fluctuations at ambient temperature (approximately kT = 25 meV), otherwise decoherence would quickly overwhelm any coherent behaviour. One such collective coherent behaviour is the superextensive charging of the Dicke quantum battery, a.k.a. superabsorption, which has been demonstrated in two experiments which we discuss here.

*Organic microcavities.* Superabsorption was first demonstrated at room temperature using a collection of molecular electronic transitions in an organic semiconductor coupled to a confined optical model in a microcavity[9]. The configuration comprises a microcavity constructed by combining two dielectric mirrors and incorporating a thin coating of a lightweight molecular semiconductor dispersed within a polymer matrix (Fig. 2a). The particular organic semiconductor employed in this investigation was the Lumogen-F Orange (LFO) dye. By operating in the vicinity of the 0-0 electronic transition, which is the transition between the lowest vibrational states of the ground and excited electronic states, the LFO molecules can be regarded as two-level systems. Through their interaction with a shared cavity mode, the conducted experiment was interpreted with the Dicke Hamiltonian. The number of coupled two-level systems was controlled by regulating the concentrations of the dye molecules. The charging and energy storage dynamics were characterized using ultrafast transient-absorption spectroscopy. In this technique, the LFO molecules in the microcavity were excited with an ultrashort pump pulse, and the stored energy as a function of time was measured with a second delayed ultrashort probe pulse, allowing femtosecond charging times to be measured. The experimental data indicated superextensive energy storage capacity and charging. This pioneering experiment demonstrated superextensive charging, but not yet the ability to store the absorbed light energy, which was emitted by the cavity on an ultrafast timescale. Future work will implement energy storage capabilities in such systems by transferring the light energy absorbed by the dye molecules to long-lived metastable states, via either energy transfer or charge separation processes. This promising work also opens up the possibility for the quantum enhancement of absorption in solar cells.

*Nuclear spins.* Another approach involves the utilization of nuclear magnetic resonance to investigate the injection and extraction of energy in nuclear spin systems[10]. The researchers examined molecular structures that exhibited a star-like configuration, comprising a designated *battery-spin* encompassed by a range of *charger-spins* numbering from 3 to 36 (Fig. 2b). In this study, the battery-spin was represented by a solitary spin-1/2 atom, while the charger-spins comprised a collection of *N* spin-1/2 atoms. To disrupt the initial thermal equilibrium state of the charger-spins, their populations were inverted using a $\pi$-pulse, which promotes all the spins from the ground to the excited state. Subsequently, the energy of the battery-spin was measured over time, taking into account the number *N* of charger-spins. Quantum state tomography, i.e. full reconstruction of a quantum state by a series of ensemble measurements, was employed to identify a charging power advantage proportional to $\sqrt{N}$. To maintain the battery-spin in a charged state, the charger-spins were consistently recharged (i.e., driven out of equilibrium) after a delay. Notably, the researchers also introduced a *load-spin*, where the battery-spin could transfer its energy after an appropriate storage period. In the complete 38-spin system, the battery-spin exhibited the ability to store energy for a maximum duration of 2 minutes. This outcome represents an encouraging advancement towards the realization of quantum batteries operating at room temperature and based on nuclear spins.

*Cold-temperature experiments*

Although the operating conditions of cold-temperature quantum batteries are more restrictive than their room-temperature counterparts, low decoherence levels allows for more subtle quantum mechanical behaviour to manifest. Here we discuss several experiments with superconductors and quantum dots that have been used to probe the properties of quantum batteries at low temperatures.

*Superconductors.* Superconductors are materials that exhibit zero electrical resistance when cooled below a critical temperature. This property is crucial for various applications, including the development of quantum computers, where superconducting circuits are used to encode and manipulate quantum information through the controlled flow of supercurrents and the creation of qubits. Hu et al.[11] conducted an experimental demonstration of a prototype for a superconducting quantum battery. In their setup, they utilized a single-mode cavity connected to a superconducting qutrit, which is a three-level variation of a qubit system (Fig 2c). The qutrit utilizes the three lowest energy levels of a transmon, which is a peculiar type of superconducting charge qubit. Since their configuration involved only a single quantum unit, it did not exhibit any collective advantage. Nonetheless, the work by Hu et al. represents a crucial initial stride towards the development of highly controllable superconducting quantum devices designed for energy storage. The researchers achieved this by employing time-dependent Rabi frequencies (i.e. the frequencies at which the population difference of two energetic levels excited by an electromagnetic field oscillates) in two microwave pulses to resonantly drive the qutrit, enabling the implementation of various controlled adiabatic charging processes. Additionally, they successfully reconstructed the complete density matrix of the qutrit through quantum state tomography, allowing them to accurately measure the stored energy at each stage of the charging and discharging dynamics. Despite the ultralow operating temperature (30 mK for the experiment by Hu et al.), the superconducting quantum battery may find promising applications in combination with superconducting quantum computers, which also operate at such ultralow temperatures, providing energy to their logic gates in a continuous and reversible fashion.

*Quantum dots.* Quantum dots are tiny semiconductor nanoparticles, typically ranging in size from 2 to 10 nanometers, that exhibit quantum mechanical effects. They can confine and control the behaviour of electrons, leading to unique optical and electrical properties, making them useful in applications such as display technologies, solar cells, and biological imaging. Wenniger et al. [12] conducted an experimental investigation into the exchange of energy between a quantum dot and an electromagnetic mode reservoir (Fig. 2d). Their study focused on the transfer of energy from an InGaAs quantum dot (referred to as the charger) to a micropillar optical cavity (referred to as the battery) through spontaneous emission. To induce the desired effects, the quantum dot was excited using a pulsed Ti:Sapphire laser in a cryostat set at temperatures ranging from 5-20 K, resulting in the dot being in a coherent superposition of the ground and excited states. The researchers also considered a phase involving work extraction, where the energy stored in the battery was transferred to another system consisting of a laser field, utilizing homodyne-type interference.

*Quantum computers.* Due to the deep link between information and energy, any platform that can act as a quantum computer could in principle, also act as a quantum battery. This was recently demonstrated on the IBM Q quantum computing platform[13] (Fig 2e). The authors of this study examined the efficiency of a qubit in terms of energy storage and charging time, which is driven by a pulse instruction. They demonstrated that by selecting a moderately wide and rapidly decaying classical driving field, it is possible to achieve excellent energy storage (surpassing 95%) in a remarkably short period (less than 135 ns), relative to typical relaxation and dephasing times. Interestingly, whilst errors in state initialisation are considered detrimental for quantum computers, the authors showed that in an energy context these errors actually enhanced the performance of the quantum battery.

**Outlook**

There is little doubt that the demand for energy storage in the future will only increase. This demand will need to be solved with novel ideas – quantum batteries offer one such solution. Although the technology is still in its infancy, it is becoming increasingly clear that idea of quantum batteries should be divided into two camps: room and cold-temperature quantum batteries. Because room-temperature quantum batteries are less reliant on the fragile nature of quantum states, they are more likely to be scaled-up and operate at energy scales and environments that will allow them to power conventional devices. Cold-temperature quantum batteries on the other hand are highly sensitive to decoherence, and therefore are unlikely to be scaled-up to power conventional devices. Instead, their likely realm of applicability will be to interface with other quantum technologies. For example, for quantum computers to perform truly reversible operations, these gates will need to be powered by quantum batteries[14]. This suggests that quantum batteries will be an integral component in future quantum computer designs. So far, quantum battery research has been focussed on the charging and discharging of energy. However, utilising quantum mechanics to improve energy density is also an important aspect of quantum batteries. Ideas have been proposed, including storing energy in the nuclei excitations[15] and nanovacuum tubes[16]. Quantum batteries are a part of the broader field of quantum energy, which investigates the role that quantum mechanics plays in the conversion, storage, and transport of energy; it provides a glimpse into a new vista in quantum-driven solutions to future energy challenges and opportunities.

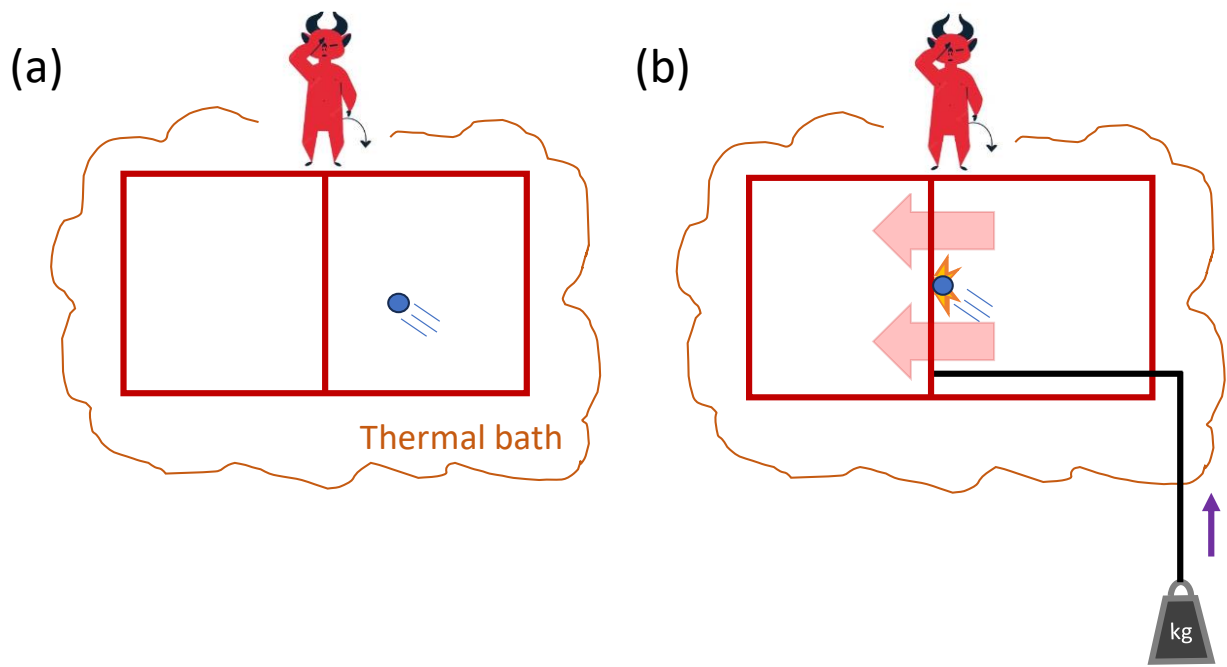

**Figure 1**. In this simplified version of a thought experiment known as *Maxwell's demon*, there is particle confined to a box in a thermal bath. (a) The demon observes that the particle is on the left side of partition. (b) Using this bit of information, he attaches a weight to the left side of the partition, so that through an adiabatic expansion process, the weight is lifted. In doing so, the demon has converted one bit of information to work.

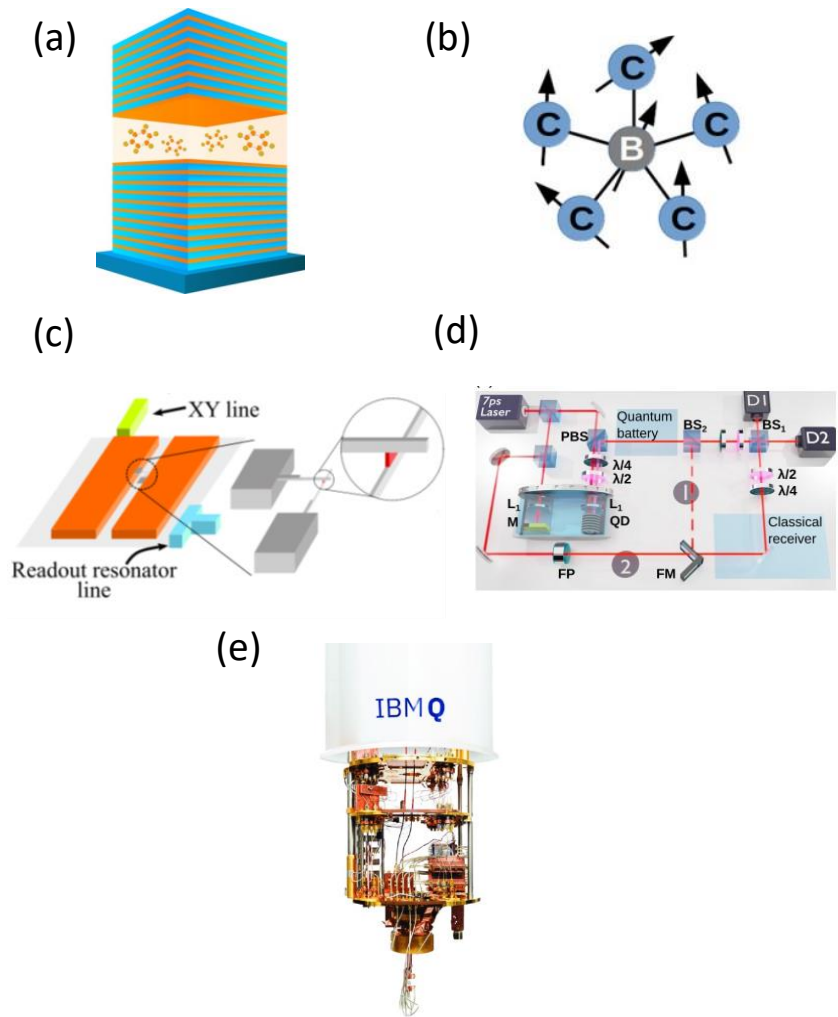

**Figure 2**. Experimental quantum battery platforms. (a) Ensembles of organic semiconducting molecules coupled to a confined optical mode in a microcavity have demonstrated superextensive charging[9]. (b) The injection and extraction of energy in molecular structures comprising of a designated *battery-spin* surrounded by a range of *charger-spins* have been investigated with nuclear magnetic resonance techniques[10]. (c) A single-mode cavity connected to a superconducting qutrit has been designed for highly controllable energy storage [11]. (d) The exchange of energy between a quantum dot and an electromagnetic mode reservoir have been investigated using interferometric setups[12]. (e) Quantum computers have been used as an quantum energy storage platform, demonstrating the deep link between information and energy storage [13].

**Author biographies:**

**James Quach** is a Science Leader at the CSIRO (Commonwealth Scientific and Industrial Research Organisation), where he leads the Quantum Batteries team. He is the inaugural Chair of the International Conference on Quantum Energy. Previously he was a Ramsay Fellow at The University of Adelaide, a Marie Curie Fellow at the Institute of Photonics Science in Barcelona, and a JSPS Fellow at the University of Tokyo. He completed his PhD at the University of Melbourne in Physics. His research interest is in all things quantum, working in quantum technology, quantum computing, quantum biology, quantum chaos, quantum thermodynamics, and quantum gravity.

**Giulio Cerullo** is a Full Professor with the Physics Department, Politecnico di Milano, where he leads the Ultrafast Optical Spectroscopy laboratory. Prof. Cerullo's research activity deals with the generation of tunable few-optical-cycle light pulses and their application to the study of ultrafast processes in (bio)-molecules and quantum confined solids. He has been General Chair of CLEO/Europe and Ultrafast Phenomena conferences. He is a Fellow of the Optical Society and of the European Physical Society and a member of Accademia dei Lincei. In 2023 he received the Quantum Electronics Prize from the European Physical Society.

**Tersilla Virgili** is a senior researcher at the Institute of Photonics and Nanotechnologies of the National Research Council (CNR) in Milano, Italy. She graduated in Physics at the University of Bologna and got her PhD at the University of Sheffield (UK). Her scientific activity is represented by more than 90 publications international and different book contributions. Her scientific interest is mainly based on the following areas: 1) spectroscopy of organic material, hybrid and inorganic for photonic and photovoltaic devices; 2) photophysics of organic microcavities working in weak and strong coupling regime; 3) organic quantum batteries.